\documentclass[twocolumn,showpacs,reprint,amsmath,amssymb,aps,prab]{revtex4-2}

\usepackage{graphicx}
\usepackage{dcolumn}
\usepackage{bm}
\usepackage{hyperref}
\hypersetup{colorlinks=true, citecolor=blue, urlcolor=blue, linkcolor=blue}

\newcommand{\be}{\begin{equation}}
\newcommand{\ee}{\end{equation}}
\newcommand{\bea}{\begin{eqnarray}}
\newcommand{\eea}{\end{eqnarray}}

\newcommand{\Angstrom}{{\buildrel _{\circ} \over {\mathrm{A}}}}

\begin{document}


\title{Feasibility study for the hard x-ray free electron laser based on synergistic use of conventional and plasma accelerator technologies}

\author{Nikolai Yampolsky}
\affiliation{Los Alamos National Laboratory, Los Alamos, New Mexico, 87545, USA}

\author{Sandra Biedron}
\affiliation{University of New Mexico, Albuquerque, New Mexico, 87131, USA}

\author{Bjorn Manuel Hegelich}
\affiliation{University of Texas, Austin, Texas, 78712, USA}

\author{Scott Luedtke}
\affiliation{Los Alamos National Laboratory, Los Alamos, New Mexico, 87545, USA}

\author{Evgenya Simakov}
\affiliation{Los Alamos National Laboratory, Los Alamos, New Mexico, 87545, USA}

\author{Stephen Milton}
\affiliation{Los Alamos National Laboratory, Los Alamos, New Mexico, 87545, USA}

	
\begin{abstract}
We access the possibility of using a conventional RF accelerator as an injector for the plasma driven wakefield accelerator. Conventional accelerators deliver high quality beams with low emittance and low energy spread. Once injected into the plasma wake, the emittance may be preserved upon proper beam matching while the energy spread may not due to long beam duration delivered by the conventional accelerator. Parameters of the overall accelerator system and the free electron laser which uses such a beam are estimated.
\end{abstract}

\maketitle

\section{Introduction}
\label{sec:Intro}
Hard x-ray free electron lasers (XFELs) are great diagnostic tools in modern science. A number of XFELs have been constructed worldwide in the past decade \cite{LCLS, Spring-8, XFEL, PAL-XFEL, SWISSFEL}. In the past, Los Alamos National Laboratory has expressed interest in hosting a similar facility to support ongoing Programs \cite{MaRIE}. The facility has been envisioned based on a conventional radio frequency (RF) accelerator technology. As a result, the estimated size and cost of the anticipated facility turned out to be large, which significantly decreased attractiveness of this project. 

The cost and size of an XFEL are driven mostly by the cost and size of the electron linear accelerator (linac). The electron beam in an XFEL must reach the energy above 10 GeV in order to generate radiation with wavelength below 1 $\Angstrom$ with conventional undulators. The required beam energy translates into about 1 km-long accelerator built with conventional RF technology. Advanced accelerator concepts are currently under development and may provide solution for increasing accelerating gradients by several orders of magnitude \cite{LWFA, PWFA, DWA, DLA}. So far, the most mature concept is acceleration of particles in a plasma wave \cite{LWFA, PWFA}.

Programs for developing FELs using the laser wakefield accelerator (LWFA) technology exist all over the world \cite{LFEL-Germany, LFEL-ELI, LFEL-Japan, LFEL-France, LFEL-BELLA, EuPraxia, LFEL-China}. So far, there has been a single experimental demonstration of this concept at soft x-ray wavelengths \cite{LFEL-China}. The main challenge in adopting the LWFA technology for FELs is yet insufficient quality of the accelerated beams \cite{LFEL-review1, LFEL-review2}. The accelerated electrons in LWFAs typically originate from the background plasma \cite{injection1, injection2, injection3, injection4, injection5, injection6}. That naturally does not allow for much control over injection. As a result, injection typically happens over an extended distance rather than instantaneously and the resulting electron bunch has large energy spread.

Alternatively, plasma accelerators can accelerate electron beams produced and pre-accelerated by conventional RF accelerators \cite{injection_aspects}. The use of the RF accelerator as an injector to an LWFA is motivated by the high quality of the injected beam compared to beams produced inside plasma during self-injection. Such a scheme would take advantage of high quality beams produced  in RF accelerators and fast acceleration to full energy in a strong field provided by plasma. In principle, this combination may produce high quality electron beams at high energy required for XFELs.
The purpose of this paper is to explore possibility of utilizing plasma accelerator technology for XFELs. We review the current state of the art and explain challenges in achieving high enough maturation of the technology.

\section{Pre-conceptual design of an FEL based on laser accelerator technology}

In this section we put together a sample pre-conceptual design of the LWFA based on an RF injector, which is suitable for the XFELs. We focus on designing a system, which closely matches the XFEL designed for MaRIE \cite{MaRIE_design1, MaRIE_design2}.

The goal is to achieve FEL lasing at photon energy of 42 keV. We assume the use of the undulator, which has already been designed for MaRIE \cite{MaRIE_design2}. That requires the electron beam to reach the energy of 12 GeV and peak current of 3 kA at the undulator entrance. Most of the acceleration is envisioned to happen in plasma. The uncertainty in the beam's quality and its high energy after the LWFA stage does not allow for temporal compression of the beam prior to undulator. That implies that the electron beam needs to be fully compressed prior to injection into the LWFA. This setup allows to keep the induced energy spread at the minimum level since the bunch inside the plasma wake is short. Reaching beam current of 3 kA is a challenging task for high brightness accelerators. The MaRIE linac design was able to reach high current at energies as low as 1 GeV \cite{MaRIE_design1}. It is extremely difficult to design an accelerator reaching full compression at significantly lower energy. As a result, the RF accelerator must provide acceleration for high brightness beams to the energy of $\sim$1 GeV with the peak current of $\sim$ 3 kA. The conventional approach to designing such systems is to implement 2-3 compressors at intermediate energies \cite{LCLS, Spring-8, XFEL, PAL-XFEL, SWISSFEL, MaRIE_design1}. In this paper we assume that a low energy linac section designed for MaRIE can be used to deliver the beam with required parameters.

\begin{widetext}
\onecolumngrid
\begin{figure}[ht]
	\center
	\includegraphics[width=1\textwidth]{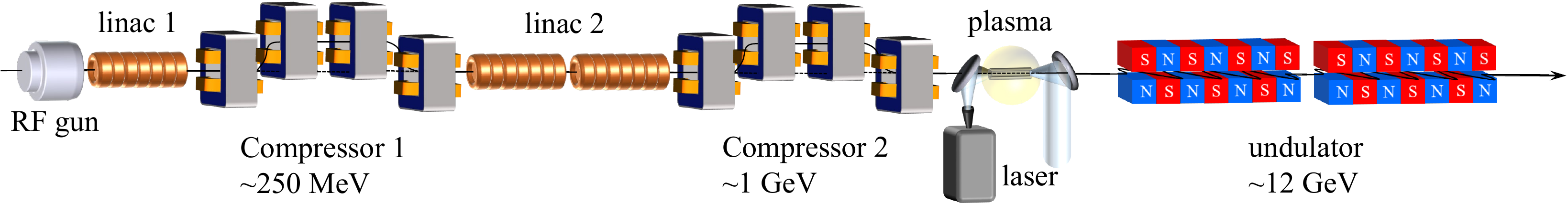}
	\caption{Schematics of the free electron laser based on the plasma accelerator technology and RF linac as injector}
	\label{fig:schematics}
\end{figure}
\twocolumngrid
\end{widetext}

The schematics of an XFEL based on the LWFA technology and an RF injector is shown in Fig.~\ref{fig:schematics}.

\begin{table}[ht]
\begin{ruledtabular}
\caption{Key design parameters for XFEL based on a LWFA}
\begin{tabular}{lll}
{\bf Laser parameters} & &\\
Laser wavelength 	& 	1 	& $\mu$m\\
Laser duration		&	97	&	fs\\
Laser radius		& 	50	&	$\mu$m\\
Laser intensity		&	$5.36\cdot 10^{18}$	&	W/cm$^2$\\
Relativistic amplitude	& 1.4 &\\
Laser pulse energy	&	41	& J\\
{\bf Plasma parameters} & &\\
Plasma density		&	$1.33\cdot 10^{17}$	& cm$^{-3}$\\
Plasma wavelength	&	100	&	$\mu$m\\
Plasma length		&	54	&	cm\\
{\bf Injected beam parameters} & &\\
Bunch energy				&	1	& GeV\\
Bunch current				&	3000& A\\
Bunch length		&	1	&	$\mu$m\\
Bunch radius		&	0.31	&	$\mu$m\\
Bunch charge		&	10	&	pC\\
Normalized emittance	&	0.2	& $\mu$m\\
Beta function		&	0.91&	mm\\
Slice energy spread &	1.8	&	MeV\\
Rms energy spread &	1.8	&	MeV\\
{\bf Extracted beam parameters} & &\\
Bunch energy				&	12	& GeV\\
Bunch current				&	3000& A\\
Bunch length		&	1	&	$\mu$m\\
Bunch radius		&	0.16	&	$\mu$m\\
Bunch charge		&	10	&	pC\\
Normalized emittance	&	0.2	& $\mu$m\\
Beta function		&	3.38&	mm\\
Slice energy spread &	3.3	&	MeV\\
Rms energy spread 	&	835	&	MeV\\
{\bf FEL parameters} & &\\
Undulator wavelength	&	1.86	&	cm\\
Undulator strength 		&	1.22	&\\
Undulator length		&	87	& m\\
Photon energy		&	42	& keV\\
Radiation wavelength	& 0.3	& $\Angstrom$\\
3D gain length		&	4.33	& m\\
Radiation radius	&	11.6	&	$\mu$m\\
Radiation bandwidth	&	14	&	\%\\
Radiation power		&	5.6	&	GW\\
Number of photons	&	$2.76\cdot 10^9$	&
\end{tabular}~\label{tbl:FEL}
\end{ruledtabular}
\end{table}

The energy of the electron beam being injected into the LWFA is high. To large degree, this concept is similar to the successive acceleration in a multi-stage LWFA. There is a strong belief in the filed that efficient injection can be implemented in a quasi-linear regime of the plasma wave \cite{collider1, collider2, EuPraxia}. This regime is characterized by a weakly nonlinear excited plasma wave, in contrast to the highly nonlinear regime ({\it e.g.} bubble regime) \cite{bubble}. Weekly nonlinear regime provides high efficiency of charge coupling, it is less sensitive to distortions of the laser pulse and plasma-driven beam instabilities, and does not result in self-injection of plasma electrons exhibiting a dark current. On the other hand, the quasi-linear regime of acceleration does not provide sufficient self-focusing, so it needs to be complimented by an external laser guidance, {\it e.g.} capillary \cite{capillary} or plasma channel \cite{channel} to overcome laser diffraction and extend interaction region for enhanced efficiency.

A large number of relevant parameters for such a system can be estimated using scaling laws of LWFAs \cite{LWFA_theory} and FELs \cite{MingXie}. These parameters are listed in Table~\ref{tbl:FEL}. The parameters are estimated self-consistently balancing multiple constraints at different stages of the machine.
Order-of-magnitude estimates based on parameters summarized in Table~\ref{tbl:FEL} show that it is possible to design an FEL at hard x-ray wavelengths, which is based on the laser accelerator technology. External injection of electrons into the plasma accelerator results in a beam of a high enough quality to achieve lasing.

\section{Key technological gaps}

Possibility of injecting external beams into a plasma wakefield has been demonstrated experimentally \cite{staging1, staging2, RF_LWFA}. However, presently these experiments are at the proof-of-principle stage rather than being a well established technique. Injecting an electron beam into a plasma wake is a complicated problem, which requires advancements in several areas before the technology reaches maturity. Some of these challenges such as suppression of instabilities, increasing captured bunch charge, increasing wall plug efficiency, reduction in degradation of the beam's quality during acceleration, increasing accelerated beam's energy, {\it etc.} are general problems for the plasma-based accelerators \cite{LWFA-pespectives} and a plan for retiring those risks has been outlined \cite{LWFA-roadmap}. Combining RF and plasma accelerator technologies imposes additional complications.

\subsection{An injector for ultra-short bunches}
Electron bunches which are suitable for plasma-based accelerators must have low bunch charge on the order of few picocoulombs. Such a bunch charge is an order of magnitude smaller than conventionally produced in RF photoinjectors developed for modern light sources \cite{RFguns}. These photoinjectors can operate at such small bunch charges \cite{LCLSgun, Swissgun} but they are not optimized for this regime. As such, a dedicated research in designing guns operating at small charges \cite{Xbandgun, LWFAgun} as well as development of the beam manipulation schemes capable of producing short bunches \cite{MaRIE_design2} must be conducted.

At the same time, the use ofan  RF accelerator as an injector is not necessarily a superior option compared to  self-injection of electrons from plasma. The quality of  accelerated beam in the considered pre-conceptual design is comparable to the quality of beams achieved in  self-injection experiments \cite{highquality1,highquality2,highquality3,highquality4,highquality5,highquality6}. The main reason for the RF accelerators failing to outperform plasma-based self-injectors is due to already mentioned difficulty to generate ultra-high peak current electron bunches, which are routinely generated from plasmas. The RF injectors have to generate longer bunches in order to produce the same overall bunch charge. That translates into the increased rms energy spread acquired by the bunch during successive acceleration in a highly nonuniform plasma wakefield.

\subsection{An interface between conventional and plasma accelerator}
Matching an electron beam from a conventional RF accelerator into a plasma accelerator is a complicated task, which has only been accomplished at the proof-of-principle level \cite{RF_LWFA}. First of all, there are tight requirements on the positioning (few $\mu$m) and time jitter ($\sim$1 fs) of the beam \cite{injection_aspects}. In addition, the electron beam must be tightly focused into plasma in order to avoid transverse overfocusing leading to the degradation in the beam quality \cite{overfocusing}. The estimated beta function for the beam (similar to Rayleigh length in optics) is at the sub-mm level as indicated in Table~\ref{tbl:FEL}. Such tight focusing cannot be achieved with conventional accelerator elements used to control the beam since they are physically large objects. The use of a tapered plasma density profile as a way to overcome this issue has been actively studied \cite{plasma_profile1, plasma_profile2, plasma_profile3} but has not been demonstrated yet.

Extraction of the plasma accelerated beam and matching it to the following conventional accelerator beamline is a challenging task as well \cite{LWFA-controll}. Large energy spread in accelerated beams makes it difficult to steer them with conventional beamline elements based on magnets since their tuning depends on the beam energy. Alternative methods for controlling the beam using plasma lenses have been proposed but this technology are still at the R{\&}D stage \cite{plasma_lens1}, although rapidly reaching maturation \cite{plasma_lens2,plasma_lens3}.

\subsection{Preservation of beam quality in a LWFA}
Electron beams in LWFAs are accelerated with electro-magnetic fields which have large gradients and in the environment, which is specifically chosen to produce large collective response to external forces. Lack of high quality beams accelerated in plasma has been a long standing problem in the field \cite{LWFA-pespectives, LWFA-roadmap}. Degradation of beams coming from an RF injector is likely to be a larger problem compared to self-injection. The quality of externally injected beams is expected to be better by design. That requires  tighter beam focusing to provide proper matching inside the wake and larger local electron density inside the beam, which in turn triggers larger response from the background plasma. At the same time, it is harder to preserve high quality of bright beams compared to their less bright plasma generated counterparts.

Diagnostics of beams in LWFAs is significantly underdeveloped  \cite{LWFA_diagnostics} compared to the conventional RF accelerators. Ultra-short duration of the accelerated bunches and poor stability of plasma accelerators make many conventional techniques unsuitable. Time resolved diagnostics of the bunches is almost nonexistent with rare exceptions \cite{Lumpkin}. At the same time, lasing inside the undulator is sensitive to the slice properties of the beam rather than the rms beam quality. Development of appropriate beam diagnostics is a critical element for building, commissioning, optimizing, and operating an XFEL based on the plasma accelerator technology. 

High quality of the beams needs to be preserved inside a meter-long plasma as indicated in Table~\ref{tbl:FEL}. Creating a uniform focusing channel of such a long length and preserving laser beam quality is a challenge. Dielectric capillaries \cite{LWFA_capillary} and plasma channels \cite{long_channel} are actively used for the LWFA but the technology is not yet mature enough to meet the requirements for the XFEL.

\subsection{Lasing with electron bunches produced in LWFAs}
An XFEL based on a plasma accelerated electron beam is expected to lase and reach saturation within 100 meters as indicated in Table~\ref{tbl:FEL}. The rms energy spread of the beam is expected to be much larger compared to the energy spread of bunches delivered by  conventional RF linacs. However, this energy spread is in fact an energy chirp (energy of particles linearly changes along the bunch) rather than uncorrelated energy spread (similar to temperature in the co-moving frame). In that respect, the LWFA-driven electron bunches have unique properties, and a lot of studies can be done toward optimizing performance of an XFEL driven by a chirped beam. These studies may include the use of transverse gradient undulators (TGUs) \cite{TGU} to reduce the bandwidth of the light and beam manipulation schemes to either produce ultra-short (well below femtosecond) \cite{atosecond} or two-color \cite{2color} pulses.

\subsection{Designing experiments for FELs with unique light properties}
The light generated in XFELs driven by the plasma-based accelerators will have unique properties very different from those of XFELs driven by conventional RF linacs. In the baseline scenario, the light pulse will have a small number of photons and a large chirp. This kind of light is not suitable for imaging \cite{CXI} and coherent spectroscopy \cite{XCS} experiments, which require monochromatic light pulses. At the same time, unique properties of the light may be utilized for designing different class of experiments. The light pulse with a strong chirp opens possibility for its further compression below the attosecond duration similar to the chirped pulse amplification (CPA) technique for visible light \cite{CPA}. Strong chirp also allows for time resolved measurements with subfemtosecond resolution since wavelength sensitive streaking is equivalent to time streaking \cite{streak}.

\section{Conclusions}

At this point, the use of an RF linear accelerator as an injector for the plasma wakefield accelerator seems possible for driving x-ray free electron lasers at sub-angstrom wavelengths. Pre-conceptual design parameters of such system are listed in Table~\ref{tbl:FEL}. These parameters are derived self-consistently using scaling laws for LWFAs and FELs. There is a lot of room for improvements there and it is likely that an XFEL with better characteristics can be constructed after detailed studies are performed. At this time, plasma accelerator technology is not mature enough to be immediately implemented. A number of key components still need to be developed and demonstrated for a functional XFEL.



\begin{thebibliography}{8}


\bibitem{LCLS}
P.~Emma, R.~Akre, J.~Arthur, R.~Bionta, C.~Bostedt, J.~Bozek, A.~Brachmann, P.~Bucksbaum, R.~Coffee, F.-J.~Decker {\it et al.}, 
``{\it First lasing and operation of an angstrom-wavelength free-electron laser}'',
Nat. Photonics {\bf 4}, 641 (2010).

\bibitem{Spring-8}
T.~Ishikawa, H.~Aoyagi, T.~Asaka, Y.~Asano, N.~Azumi, T.~Bizen, H.~Ego, K.~Fukami, T.~Fukui, Y.~Furukawa {\it et al.}, ``{\it A compact X-ray free-electron laser emitting in the sub-angstrom region}'', Nat. Photonics {\bf 6}, 540 (2012).

\bibitem{XFEL}
T.~Tschentscher, C.~Bressler, J.~Grünert, A.~Madsen, A.~Mancuso, M.~Meyer, A.~Scherz, H.~Sinn, and U.~Zastrau, ``{\it Photon beam transport and scientific instruments at the European XFEL}'', Appl. Sci. {\bf 7}, 592 (2017).

\bibitem{PAL-XFEL}
I.~Ko, H.-S.~Kang, H.~Heo, C.~Kim, G.~Kim, C.-K.~Min, H.~Yang, S.~Baek, H.-J.~Choi, G.~Mun {\it et al.}, ``{\it Construction and Commissioning of PAL-XFEL Facility }'', Appl. Sci. {\bf 7}, 479 (2017).

\bibitem{SWISSFEL}
C.~J.~Milne, T.~Schietinger, M.~Aiba, A.~Alarcon  J.~Alex, A.~Anghel, V.~Arsov, C.~Beard, P.~Beaud, S.~Bettoni {\it et al.}, ``{\it SwissFEL: The Swiss X-ray Free Electron Laser}'', Appl. Sci {\bf 7}, 720 (2017).

\bibitem{MaRIE}
B.~E.~Carlsten and R.~L.~Sheffield, ``{\it 
Pre-Conceptual Design Requirements For The MaRIE Facility At LANL And The Resulting X-Ray Free Electron Laser Baseline Design}'',  Conf. Proc. C {\bf 110328}, 2417 (2011).

\bibitem{LWFA}
T.~Tajima and J.~M.~Dawson, ``{\it Laser Electron Accelerator}'', Phys. Rev. Lett. {\bf 43}, 267 (1979).

\bibitem{PWFA}
P.~Chen, J.~M.~Dawson, R.~W.~Huff, and T.~Katsouleas, ``{\it Acceleration of electrons by the interaction of a bunched electron beam with a plasma}'',  Phys.  Rev.  Lett. {\bf 54}, 693 (1985).

\bibitem{DWA}
W.~Gai, P.~Schoessow, B.~Cole, R.~Konecny, J.~Norem, J.~Rosenzweig, and J.~Simpson, ``{\it Experimental Demonstration of Wake-Field Effects in Dielectric Structures}'', Phys. Rev. Lett. {\it 61}, 2756 (1988).

\bibitem{DLA}
Xintian Eddie Lin, ``{\it Photonic band gap fiber accelerator}'', Phys. Rev. ST Accel. Beams {\bf 4}, 051301 (2001).

\bibitem{LFEL-Germany}
F.~Grüner, S.~Becker, U.~Schramm, T.~Eichner, M.~Fuchs, R.~Weingartner, D.~Habs, J.~Meyer-ter-Vehn, M.~Geissler, M.~Ferrario, {\it et al.}, ``{\it Design considerations for table-top, laser-based VUV and X-ray free electron lasers}'', Appl. Phys. B {\bf 86}, 431 (2007).

\bibitem{LFEL-ELI}
M.~M.~Aleonard {\it et al.}, ``{\it ELI Whitebook}'', Berlin, Germany: THOSS Media GmbH (2011).

\bibitem{LFEL-Japan}
Kazuhisa Nakajima, ``{\it Conceptual designs of a laser plasma accelerator-basedEUV-FEL and an all-optical Gamma-beam source}'', High Power Laser Sci. Eng. {\bf 2} e31 (2014).

\bibitem{LFEL-France}
M.~E.~Couprie, M.~Labat, C.~Evain, F.~Marteau, F.~Briquez, M.~Khojoyan, C.~Benabderrahmane, L.~Chapuis, N.~Hubert, C.~Bourassin-Bouchet, {\it et al.}, ``{\it An application of laser–plasma acceleration: towards a free-electron laser amplification}'',  Plasma Phys. Control. Fusion {\bf 58} 034020 (2016).

\bibitem{LFEL-BELLA}
J.~van~Tilborg, S.~K.~Barber, F.~Isono, C.~B.~Schroeder, E.~Esarey, and W.~P.~Leemans, ``{\it Free-electron lasers driven by laser plasma accelerators}'', AIP Conference Proceedings, {\bf 1812}, 020002 (2017).

\bibitem{EuPraxia}
R.~W.~Assmann, M.~K.~Weikum, {\it et al.}, ``{\it EuPRAXIA Conceptual Design Report}'', Eur. Phys. J. Special Topics {\bf 229}, 3675 (2020).

\bibitem{LFEL-China}
Wentao~Wang, Ke~Feng, Lintong~Ke, Changhai~Yu, Yi~Xu, Rong~Qi, Yu~Chen, Zhiyong~Qin, Zhijun~Zhang, Ming~Fang, {\it et al.}, ``{\it Free-electron lasing at 27 nanometres based on a laser wakefield accelerator}'', Nature {\bf 595}, 516 (2021).

\bibitem{LFEL-review1}
S.~Corde, K.~Ta~Phuoc, G.~Lambert, R.~Fitour, V.~Malka, A.~Rousse, A.~Beck, and E.~Lefebvre, ``{\it Femtosecond x-rays from laser-plasma accelerators}'', Rev. Mod. Phys. {\bf 85}, 1 (2013).

\bibitem{LFEL-review2}
Marie~Emmanuelle~Couprie, ``{\it Towards compact Free Electron–Laser based on laser plasma accelerators}'', Nuclear Inst. and Methods in Physics Research, {\bf A 909}, 5 (2018).

\bibitem{injection1}
E.~Esarey, R.~F.~Hubbard, W.~P.~Leemans, A.~Ting, P.~Sprangle, ``{\it Electron injection in to plasma wakefields by colliding laser pulses}'', Phys. Rev. Lett. {\bf 79}, 2682 (1997).

\bibitem{injection2}
C.~McGuffey, A.~G.~R.~Thomas, W.~Schumaker, T.~Matsuoka, V.~Chvykov, F.~J.~Dollar, G.~Kalintchenko, V.~Yanovsky, A.~Maksimchuk, K. Krushelnick, {\it et al.}, ``{\it Ionization Induced Trapping in a Laser Wakefield Accelerator}'', Phys. Rev. Lett. {\bf 104}, 025004 (2010).

\bibitem{injection3}
C.~E.~Clayton, J.~E.~Ralph, F.~Albert, R.~A.~Fonseca, S.~H.~Glenzer, C.~Joshi, W.~Lu, K.~A.~Marsh, S.~F.~Martins, W.~B.~Mori, {\it et al.}, ``{\it Self-Guided Laser Wakefield Acceleration beyond 1 GeV Using Ionization-Induced Injection}'', Phys. Rev. Lett. {\bf 105}, 105003 (2010).

\bibitem{injection4}
B.~Hidding, G.~Pretzler, J.~B.~Rosenzweig, T.~Königstein, D.~Schiller, and D.~L.~Bruhwiler, ``{\it  Ultracold electron bunch generation via plasma photocathode emission and acceleration in a beam-driven  plasma  blowout}'',  Phys.  Rev.  Lett.  {\bf 108},  035001 (2012).  

\bibitem{injection5}
R.~Lehe, A.~F.~Lifschitz, X.~Davoine, C.~Thaury, and V.~Malka, ``{\it Optical Transverse Injection in Laser-Plasma Acceleration}'', Phys. Rev. Lett. {\bf 111}, 085005 (2013).

\bibitem{injection6}
A.~Buck, J.~Wenz, J.~Xu, K.~Khrennikov, K.~Schmid, M.~Heigoldt, J.~M.~Mikhailova, M.~Geissler, B.~Shen, F.~Krausz {\it et al.}, ``{\it Shock-Front Injector for High-Quality Laser-Plasma Acceleration}'', Phys. Rev. Lett. {\bf 110}, 185006 (2013).

\bibitem{injection_aspects}
Barbara~Marchetti, Ralph~Assmann, Ulrich~Dorda and Jun~Zhu, ``{\it Conceptual and Technical Design Aspects of Accelerators for External Injection in LWFA}'', Appl. Sci. {\bf 8}, 757 (2018).

\bibitem{MaRIE_design1}
J.~W.~Lewellen, K.~A.~Bishofberger, B.~E.~Carlsten, L.~D.~Duffy, F.~L.~Krawczyk, Q.~R.~Marksteiner, D.~C.~Nguyen, S.~J.~Russell, R.~L.~Sheffield and N.~Yampolsky, ``{\it Status of the MaRIE X-FEL Accelerator Design}'',In Proceedings of the 6nd International Particle Accelerator Conference IPAC2015, Richmond, VA, USA, 1894 (2015).

\bibitem{MaRIE_design2}
Bruce~E.~Carlsten, Petr~M.~Anisimov, Cris~W.~Barnes, Quinn~R.~Marksteiner, River~R.~Robles and Nikolai~Yampolsky, ``{\it High-Brightness Beam Technology Development for a Future Dynamic Mesoscale Materials Science Capability}'', Instruments {\bf 3}, 52 (2019).

\bibitem{collider1}
C.~B.~Schroeder, E.~Esarey, C.~G.~R.~Geddes, C.~Benedetti, and W.~P.~Leemans, ``{\it Physics considerations for laser-plasma linear colliders}'', Phys. Rev. ST Accel. Beams {\bf 13}, 101301 (2010).

\bibitem{collider2}
Kazuhisa~Nakajima, Aihua~Deng, Xiaomei~Zhang, Baifei~Shen, Jiansheng~Liu, Ruxin~Li, Zhizhan~Xu, Tobias~Ostermayr, Stefan~Petrovics, Constantin~Klier {\it et al.} ``{\it Operating plasma density issues on large-scale laser-plasma accelerators toward high-energy frontier}'', Phys. Rev. ST Accel. Beams 14, 091301 (2011).

\bibitem{bubble}
J.~B.~Rosenzweig, B.~Breizman, T.~Katsouleas, and J.~J.~Su, ``{\it Acceleration and focusing of electrons in two-dimensional nonlinear plasma wake fields}'', Phys. Rev. A {\bf 44}, R6189(R) (1991).

\bibitem{capillary}
A.~Zigler, Y.~Ehrlich, C.~Cohen, J.~Krall, and P.~Sprangle, ``{\it Optical guiding of high-intensity laser pulses in a long plasma channel formed by a slow capillary discharge}'', J. Opt. Soc. Am. B {\bf 13}, 68 (1996).

\bibitem{channel}
C.~G.~Durfee~III and H.~M.Milchberg, ``{\it Light pipe for high intensity laser pulses}''  Phys. Rev. Lett. {\bf 71}, 2409 (1993).

\bibitem{LWFA_theory}
E.~Esarey, C.~B.~Schroeder, and W.~P.~Leemans, ``{\it Physics of laser-driven plasma-based electron accelerators}'', Rev. Mod. Phys. {\bf 81}, 1229 (2009).

\bibitem{MingXie}
Ming Xie, ``{\it Design optimization for an X-ray free electron laser driven by the SLAC linac}'',  In Proceedings of the IEEE 1995 Particle Accelerator Conference, Dallas, TX, USA; IEEE Cat. No. 95CH35843 {\bf 3}, 183 (1995).

\bibitem{staging1}
S.~Steinke, J.~van~Tilborg, C.~Benedetti, C.~G.~R.~Geddes, C.~B.~Schroeder, J.~Daniels, K.~K.~Swanson, A.~J.~Gonsalves, K.~Nakamura, N.~H.~Matlis {\it et al.}, ``{\it Multistage coupling of independent laser-plasma accelerators}'', Nature {\bf 530}, 190 (2016).

\bibitem{staging2}
T.~Kurz, T.~Heinemann, M.~F.~Gilljohann, Y.~Y.~Chang, J.~P.~Couperus Cabadağ, A.~Debus, O.~Kononenko, R.~Pausch, S.~Schöbel, R.~W.~Assmann {\it et al.}, ``{\it Demonstration of a compact plasma accelerator powered by laser-accelerated electron beams}'', 
Nature Comm. {\bf 12}, 2895 (2021).

\bibitem{RF_LWFA}
Yipeng~Wu, Jianfei~Hua, Zheng~Zhou, Jie~Zhang, Shuang~Liu, Bo~Peng, Yu~Fang, Xiaonan~Ning, Zan~Nie, Fei~Li {\it et al.}, ``{\it High-throughput injection–acceleration of electron bunches from a linear accelerator to a laser wakefield accelerator}'', Nature Phys. {\bf 17},801 (2021).

\bibitem{LWFA-pespectives}
C.~Joshi, S.~Corde, and W.~B.~Mori, ``{\it Perspectives on the generation of electron beams from plasma-based accelerators and their near and long term applications}'', Phys. Plasmas {\bf 27}, 070602 (2020).

\bibitem{LWFA-roadmap}
Eric~R. Colby and L.~K.~Len, ``{\it Roadmap to the Future}'', Rev. Accel. Sci. Tech. {\bf 9}, 1 (2016).

\bibitem{RFguns}
G.~Penco, E.~Allaria, L.~Badano, P.~Cinquegrana, P.~Craievich, M.~Danailov, A.~Demidovich, R.~Ivanov, A.~Lutman, L.~Rumiz {\it et al.}, ``{\it Optimization of a high brightness photoinjector for a seeded FEL facility}'', J. Inst. {\bf 8}, P05015 (2013).

\bibitem{LCLSgun}
Y.~Ding, A.~Brachmann, F.-J.~Decker, D.~Dowell, P.~Emma, J.~Frisch, S.~Gilevich, G.~Hays, Ph.~Hering, Z.~Huang {\it et al.}, ``{\it Measurements and Simulations of Ultralow Emittance and Ultrashort Electron Beamsin the Linac Coherent Light Source}'', Phys. Rev. Lett. {\bf 102}, 254801 (2009).

\bibitem{Swissgun}
E.~Prat, P.~Dijksta, M.~Aiba, S.~Bettoni, P.~Craievich, E.~Ferrari, R.~Ischebeck, F.~Löhl, A.~Malyzhenkov, G.~L.~Orlandi, S.~Reiche, and T.~Schietinger, ``{\it Generation and Characterization of Intense Ultralow-Emittance Electron Beams forCompact X-Ray Free-Electron Lasers}'', Phys. Rev. Lett. {\bf 123}, 234801 (2019).

\bibitem{Xbandgun}
Yipeng~Sun, Chris~Adolphsen, Cecile~Limborg-Deprey, Tor~Raubenheimer, and Juhao~Wu, ``{\it Low-charge, hard x-ray free electron laser driven with an X-band injector and accelerator}'', Phys. Rev. ST Accel. Beams {\bf 15}, 030703 (2012).

\bibitem{LWFAgun}
A.~Giribono, A.~Bacci, E.~Chiadroni, A.~Cianchi, M.~Croia, M.~Ferrario, A.~Marocchino, V.~Petrillo, R.~Pompili, S.~Romeo {\it et al.}, ``{\it RF injector design studies for the trailing witness bunch for a plasma-based user facility}'', Nuclear Inst. and Methods in Physics Research, A {\bf 909}, 229 (2018).

\bibitem{highquality1}
J.~Faure, C.~Rechatin, A.~Norlin, A.~Lifschitz, Y.~Glinec, and V.~Malka, ``{\it Controlled injection and acceleration of electrons in plasma wakefields by colliding laser pulses}'', Nature {\bf 444}, 737 (2006).

\bibitem{highquality2}
J.~Faure, C.~Rechatin, O.~Lundh, L.~Ammoura, and V.~Malka, ``{\it Injection and acceleration of quasimonoenergetic relativistic electronbeams using density gradients at the edges of a plasma channel}'', Phys. Plasmas {\bf 17}, 083107 (2010).

\bibitem{highquality3}
O.~Lundh, J.~Lim, C.~Rechatin, L.~Ammoura, A.~Ben-Ismaïl, X.~Davoine, G.~Gallot, J-P.~Goddet, E.~Lefebvre, V.~Malka, and J.~Faure, ``{\it Few femtosecond, few kiloampere electron bunch produced by a laser–plasma accelerator}'', Nature Phy. {\bf 7}, 219 (2011).

\bibitem{highquality4}
B.~B.~Pollock, C.~E.~Clayton, J.~E.~Ralph, F.~Albert, A.~Davidson, L.~Divol, C.~Filip, S.~H.~Glenzer, K.~Herpoldt, W.~Lu {\it et al.}, ``{\it Demonstration of a Narrow Energy Spread, $\sim$0.5 GeV Electron Beam from a Two-Stage Laser Wakefield Accelerator}'', Phys. Rev. Lett. {\bf 107}, 045001 (2011).

\bibitem{highquality5}
R.~Weingartner, S.~Raith, A.~Popp, S.~Chou, J.~Wenz, K.~Khrennikov, M.~Heigoldt, A.~R.~Maier, N.~Kajumba, M.~Fuchs {\it et al.}, ``{\it Ultralow emittance electron beams from a laser-wakefield accelerator}'', Phys. Rev. ST Accel. Beams {\bf 15}, 111302 (2012).

\bibitem{highquality6}
W.~P.~Leemans, A.~J.~Gonsalves, H.-S.~Mao, K.~Nakamura, C.~Benedetti, C.~B.~Schroeder, Cs.~Tóth, J.~Daniels, D.~E.~Mittelberger, S.~S.~Bulanov {\it et al.}, ``'{\it Multi-GeV Electron Beams from Capillary-Discharge-Guided Subpetawatt Laser Pulses in the Self-Trapping Regime}', Phys. Rev. Lett. {\bf 113}, 245002 (2014).

\bibitem{overfocusing}
T.~Mehrling, J.~Grebenyuk, F.~S.~Tsung, K.~Floettmann, and J.~Osterhoff, ``{\it Transverse emittance growth in staged laser-wakefield acceleration}'', Phys. Rev. ST Accel. Beams {\bf 15}, 111303 (2012).

\bibitem{plasma_profile1}
Klaus~Floettmann, ``{\it Adiabatic matching section for plasma accelerated beams}'', Phys. Rev. ST Accel. Beams {\bf 17}, 054402 (1214).

\bibitem{plasma_profile2}
X.~L.~Xu, J.~F.~Hua, Y.~P.~Wu, C.~J.~Zhang, F.~Li, Y.~Wan, C.-H.~Pai, W.~Lu, W.~An, P.~Yu {\it et al.}, ``{\it Physics of Phase Space Matching for Staging Plasma and Traditional Accelerator Components Using Longitudinally Tailored Plasma Profiles}'', Phys. Rev. Lett. {\bf 116}, 124801 (2016).

\bibitem{plasma_profile3}
E.~Svystun, R.~W.~Assmann, U.~Dorda, A.~Ferran~Pousa, T.~Heinemann, B.~Marchetti, A.~Martinez~de~la~Ossa, P.~A.~Walker, M.~K.~Weikum, and J.~Zhu, ``{\it Beam quality preservation studies in a laser-plasma accelerator with external injection for EuPRAXIA}'', Nuclear Inst. and Methods in Physics Research, A {\bf 909}, 90 (2018).

\bibitem{LWFA-controll}
T.~André, I.~A.~Andriyash, A.~Loulergue, M.~Labat, E.~Roussel, A.~Ghaith, M.~Khojoyan, C.~Thaury, M.~Valléau, F.~Briquez, {\it et al.}, ``{\it Control of laser plasma accelerated electrons for light sources}'', Nature Comm. {\bf 9}, 1334 (2018).

\bibitem{plasma_lens1}
G.~Hairapetian, P.~Davis, C.~E.~Clayton, C.~Joshi, S.~C.~Hartman, C.~Pellegrini, and T.~Katsouleas, ``{\it Experimental demonstration of dynamic focusing of a relativistic electron bunch by an overdense plasma lens}'', Phys. Rev. Lett. {\bf 72}, 2403 (1994).

\bibitem{plasma_lens2}
S.~Kuschel, D.~Hollatz, T.~Heinemann, O.~Karger, M.~B.~Schwab, D.~Ullmann, A.~Knetsch, A.~Seidel, C.~Rödel, M.~Yeung {\it et al.}, ``{\it Demonstration of passive plasma lensing of a laser wakefield accelerated electron bunch}'', Phys. Rev. Accel. Beams {\bf 19}, 071301 (2016).

\bibitem{plasma_lens3}
E.~Chiadronia, M.~P.~Ananiaa, M.~Bellavegliaa, A.~Biagionia, F.~Bisestoa, E.~Brentegania, F.~Cardellia, A.~Cianchib, G.~Costaa, D.~Di~Giovenale {\it et al.}, ``{\it Overview of plasma lens experiments and recent results at SPARC LAB}'', Nuclear Inst. and Methods in Physics Research, A {\bf 909}, 16 (2018).

\bibitem{LWFA_diagnostics}
M.~C.~Downer, R.~Zgadzaj, A.~Debus, U.~Schramm, and M.~C.~Kaluza, ``{\it Diagnostics for plasma-based electron accelerators}'', Rev. Mod. Phys. {\bf 90}, 035002 (2018).

\bibitem{Lumpkin}
A.~H.~Lumpkin, M.~LaBerge, D.~W.~Rule, R.~Zgadzaj, A.~Hannasch, O.~Zarini, B.~Bowers, A.~Irman, J.~P.~Couperus~Cabada, A.~Debus {\it et al.}, ``{\it Coherent Optical Signatures of Electron Microbunching in Laser-Driven Plasma Accelerators}'', Phys. Rev. Lett. {\bf 125}, 014801 (2020).

\bibitem{LWFA_capillary}
A.~J.~Gonsalves, K.~Nakamura, J.~Daniels, C.~Benedetti, C.~Pieronek, T.~C.~H.~de~Raadt, S.~Steinke, J.~H.~Bin, S.~S.~Bulanov, J.~van~Tilborg, ``{\it Petawatt Laser Guiding and Electron Beam Acceleration to 8 GeV in a Laser-Heated Capillary Discharge Waveguide}'', Phys. Rev. Lett. {\bf 122},084801 (2019).

\bibitem{long_channel}
B.~Miao, L.~Feder, J.~E.~Shrock, A.~Goffin, and H.~M.~Milchberg, ``{\it Optical Guiding in Meter-Scale Plasma Waveguides}'', Phys. Rev. Lett. {\bf 125}, 074801 (2020).

\bibitem{TGU}
A.~Bernhard, V.~Afonso~Rodríguez, S.~Kuschel, M.~Leier, P.~Peiffer, A.~Sävert, M.~Schwab, W.~Werner, C.~Widmann, A.~Will {\it et al.}, ``{\it Progress on experiments towards LWFA-driven transverse gradient undulator-based FELs}'', Nuclear Inst. and Methods in Physics Research, A {\bf 909}, 391(2018).

\bibitem{atosecond}
A.~Marinelli, J.~MacArthur, P.~Emma, M.~Guetg, C.~Field, D.~Kharakh, A.~A.~Lutman, Y.~Ding, and Z.~Huang, ``{\it Experimental demonstration of a single-spike hard-X-ray free-electron laser starting from noise}'', Appl. Phys. Lett. {\bf 111}, 151101 (2017).

\bibitem{2color}
A.~Marinelli, D.~Ratner, A.~A.~Lutman, J.~Turner, J.~Welch, F.-J.~Decker, H.~Loos, C.~Behrens, S.~Gilevich, A.~A.~Miahnahri {\it et al.}, ``{\it High-intensity double-pulse X-ray free-electron laser}'', Nat. Communications {\bf 6}, 6369 (2015).

\bibitem{CXI}
Mengning~Liang, Garth~J.~Williams, Marc~Messerschmidt, M.~Marvin~Seibert, Paul~A.~Montanez, Matt~Hayes, Despina~Milathianaki, Andrew~Aquila, Mark~S.~Hunter, Jason~E.~Koglin {\it et al.}, ``{\it The Coherent X-ray Imaging instrument at the Linac Coherent Light Source}'', J Synchrotron Radiat. {\bf 22}, 514 (2015).

\bibitem{XCS}
Roberto~Alonso-Mori, Chiara~Caronna, Matthieu~Chollet, Robin~Curtis, Daniel~S.~Damiani, Jim~Defever, Yiping~Feng, Daniel~L.~Flath, James~M.~Glownia, Sooheyong~Lee {\it et al.}, ``{\it The X-ray Correlation Spectroscopy instrument at the Linac Coherent Light Source}'', J Synchrotron Radiat. {\bf 22}, 508 (2015).

\bibitem{CPA}
Donna~Strickland and Gerard~Mourou, ``{\it Compression of amplified chirped optical pulses}'', Opt. Commun. {\bf 56}, 219 (1985).

\bibitem{streak}
Jinyang~Liang and Lihong~V.~Wang, ``{\it Single-shot ultrafast optical imaging}'', Optica. {\bf 5}, 1113 (2018).


\end{thebibliography}
\end{document}